**Magnetic interactions in IV-VI diluted magnetic semiconductors**

*Małgorzata Górska\*, Łukasz Kilański, Andrzej Łusakowski*

Institute of Physics, Polish Academy of Sciences, Aleja Lotników 32/46, PL-02668 Warsaw, Poland

E-mail: gorska@ifpan.edu.pl

Diluted magnetic semiconductors (DMS) are interesting because of the interplay between the electronic and magnetic subsystems. We describe selected magnetic properties of IV-VI diluted magnetic semiconductors, looking at the similarities and differences between magnetic properties of II-VI, IV-VI, and III-V DMS. We focus on the influence of the crystalline and electronic structure of the material on its magnetic properties, especially on the exchange interactions among magnetic ions. We describe methods of determination of the exchange parameters by using different experimental techniques, such as measurements of magnetic susceptibility, magnetization, and specific heat. We follow the development in the material technology from bulk crystals to thin films and nanostructures.

## 1. Introduction

Diluted magnetic semiconductors (DMS) – also referred to as semimagnetic semiconductors (see e.g. References [1, 2]) – are semiconducting alloys whose cation lattice is made up in part of substitutional magnetic atoms. They are interesting for several reasons. There is possible to observe combined properties of semiconductors and magnetic materials. Their ternary (or more than ternary) structure makes it possible to change the lattice and band parameters by changing the composition of the material. That leads to a possibility of adjusting their structural and electrical properties and investigating the mutual influence of the electronic and magnetic subsystems. The dilute distribution of magnetic ions in the cation sublattice allows us to investigate their magnetic interactions in isolated pairs or small clusters. By changing the magnetic ion content we may observe phase transitions between paramagnetic, ferromagnetic, and spin-glass or spin-glass-like phases.

In the early stage of research the influence of magnetic ions on the electronic system was investigated by means of magnetooptical and magnetotransport measurements (for a review see References [3, 4]). New effects due to the modification of the band structure by the presence of magnetic ions, especially the large spin-splitting, were observed.

Magnetic properties of DMS were investigated mostly by means of the magnetic susceptibility and magnetization measurements.

Typical magnetic dopants used to obtain DMS are transition-metal-ions such as Mn, Fe, or Co and rare-earth-ions such as Eu or Gd. Semiconductors most often applied as host semiconductors are: II-VI, IV-VI, and III-V compounds. Experiments on optical and magnetic properties of II-VI DMS – $Hg_{1-x}Mn_xTe$ – started 50 years ago.[5-7] Comprehensive reviews of first 20 years of investigations on II-VI DMS may be found in References [3, 8, 9]. II-VI DMS were the first group largely studied because they could be obtained in a wide range of compositions, in some cases even with $0 < x \leq 1$.[3,9] Investigations of magnetic properties of IV-VI DMS – $Sn_{1-x}Mn_xTe$ – also started as early as in 1970,[10,11] but IV-VI DMS were not as largely investigated because their range of compositions as bulk materials was usually of an order of $0 < x \leq 0.15$. However, they turned out to be the first DMS in which a *reversible* transition between paramagnetic and ferromagnetic phase, induced by changing the carrier concentration, was observed.[12] This effect was explained as due to the RKKY interaction.[13] Comprehensive reviews of first 10 years of investigations on IV-VI DMS may be found in References [2, 8, 9]. The strong influence of free carriers on the magnetic properties of DMS was next observed in III-V DMS, such as $Ga_{1-x}Mn_xAs$.[14] In III-V DMS, because of the high carrier concentrations, ferromagnetic Curie temperatures as high as 200 K were observed.[15] That suggested a possibility of applications based on the DMS and a new branch of semiconductor physics called spintronics.[16,17]

This review is focused primarily on the magnetic properties of IV-VI DMS. Because of the limited size it covers only few selected topics. We do not describe the magnetooptical, magnetotransport, neutron diffraction, and electron paramagnetic resonance experiments. We concentrate mostly on interactions among magnetic ions investigated by means of the magnetic susceptibility, magnetization, and specific heat measurements. We present magnetic properties of the most representative IV-VI DMS compounds, other DMS belonging to this group may be omitted. Even in this narrow scope the review is far from being complete. It is rather a walk along one path in research on the subject over last 60 years.

## 2. Exchange interactions

Let us consider the magnetic ion subsystem in a crystal. The Hamiltonian may be written in form:

$$H = H_0 - \sum_{i,j} J_{ij} S_i \cdot S_j - \sum_i g_M \mu_B S_i \cdot H,  \tag{1}$$

where $\mathcal{H}_0$ is the Hamiltonian of the host crystal in absence of the magnetic field, $J_{ij}$ are the pair exchange parameters, $S_i$, $S_j$ are the total ion spin operators, $g_M$ is the magnetic ion g-factor, $\mu_B$ is the Bohr magneton, and $H$ is the magnetic field. The second term in the Hamiltonian – the Heisenberg Hamiltonian – describes the exchange interaction between two magnetic ions, the third term describes the interaction of magnetic ions with the external magnetic field. The summation is over all magnetic ions.

In general, Hamiltonian (1) can be solved only in a very limited number of cases, e.g. for a pair of spins.

The energy of the magnetic ion pair is:

$$E_{pair} = -J[S_T(S_T + 1) - S_1(S_1 + 1) - S_2(S_2 + 1)] + g\mu_B S^z H,  \tag{2}$$

where $S_T$ is the total spin of a pair ($S_T = S_1 + S_2$), and $S^z = -S_T, \ldots., S_T$.

For $J > 0$ the lowest value of the first term in $E_{pair}$ corresponds to the parallel configuration of two spins, the highest value to the antiparallel configuration ($S_T = 0$), i.e, when $J$ is positive the interaction is ferromagnetic, when $J$ is negative the interaction is antiferromagnetic.

The exchange interaction originates from the electrostatic Coulomb forces coupling the electrons and from the Pauli principle. We may roughly divide the exchange mechanisms occurring in DMS into following groups:

1. Interactions between electrons localized on ions.

➢ Direct exchange – direct interaction between electrons in magnetic orbitals, like *d*-electrons in transition metal ions or *f*-electrons in rare-earth-ions. Since the interatomic distance between magnetic ions in DMS is usually of an order of several Å and atomic radii of the *d*-shells in transition metals and *f*-shells in rare earth ions are usually smaller than 1 Å,[18] the direct exchange does not play a significant role in DMS.

➢ Superexchange – interaction between magnetic ions via an anion. Experiments and theories have shown (see e.g. References [19–21]), that this is a dominant mechanism in DMS in which the free carrier concentration does not exceed about $1 \cdot 10^{20}$ cm$^{-3}$.

2. Interactions mediated by free carriers.

- Bloembergen-Rowland type exchange – interaction mediated by electrons excited in real or virtual interband transition.[22] This interaction depends strongly on the energy gap value. As was shown in References [19–21], this interaction contributes about 5% to the exchange parameter value in DMS.
- RKKY type exchange – interaction mediated by free carriers on the Fermi level.[13] It becomes important in degenerate DMS. In IV-VI DMS it was observed in materials based on SnTe and GeTe, due to the high carrier concentrations. The RKKY interaction turned out to be dominant exchange mechanism in III-V DMS, because Mn substitutes Ga as an acceptor and creates a high hole concentration.

In all cases the exchange interaction is a result of overlapping of the electronic orbitals. Mattis summarized it simply and clearly:[23]

*Exchange equals overlap.*

## 3. Magnetic susceptibility and magnetization

### 3.1. II-VI DMS and IV-VI DMS based on lead chalcogenides

We may see how exchange equals overlap by comparing magnetic properties of II-VI and IV-VI DMS. The pair exchange parameter, *J*, may be determined from either the magnetic susceptibility or magnetization measurements.

In the molecular field approximation the high-temperature magnetic susceptibility may be well described by the Curie-Weiss expression:

$$\chi = \frac{N_0 \bar{x} g_M^2 \mu_B^2 S(S+1)}{3 k_B (T+\theta)} + \chi_0. \tag{3}$$

Here $N_0$ is the number of cation sites, $k_B$ is the Boltzmann constant, $S$ is the spin of the magnetic ion, $\chi_0$ is the susceptibility of the host lattice, $\bar{x}$ is the effective concentration of magnetic ions, and $\theta$ is the paramagnetic Curie temperature. The expression (3) is correct only for $T \gg \theta$. The paramagnetic Curie temperature is related to the exchange parameter:

$$k_B \cdot \theta = \frac{-2}{3} S(S+1) \sum_i J_i z_i, \tag{4}$$

where $z_i$ is the number of magnetic ions in the *i*th coordination zone and the summation is over all coordination zones.

We usually make two more approximations:

1. In the virtual crystal approximation we replace the number of magnetic ions by the number of all cation sites multiplied by $\bar{x}$.

2. In most DMS the exchange parameter in the first coordination zone is an order of magnitude higher than in the second coordination zone, and the other ones are negligible. Therefore we take into consideration only the first coordination zone, i.e. the nearest neighbors.

As a result in Equation (4) $\sum_i J_i z_i$ is replaced by $\bar{x}Jz$, where $z$ is the number of nearest neighbors.

We now obtain the exchange parameter:

$$\frac{J}{k_B} = \frac{-3\theta}{2S(S+1)z\bar{x}}. \tag{5}$$

For the materials with zincblende, wurtzite, or rocksalt structure $z = 12$.

As an example of susceptibility data for II-VI and IV-VI DMS we show in Figures 1 and 2 the inverse susceptibility, $\chi^{-1}$, versus temperature.[24,25] The $\chi^{-1}$ is nearly linear with $T$; the apparent deviation from a straight line is a result of the host diamagnetic contribution. The curves extrapolate to the negative temperatures as $\chi^{-1}$ goes to zero, implying a weak antiferromagnetic interaction, $J/k_B$.

In II-VI DMS the Curie-Weiss temperatures are usually an order of magnitude higher than in the IV-VI DMS. That indicates that the values of the exchange parameter in II-VI DMS are also an order of magnitude higher than in IV-VI DMS. In the temperature dependence of magnetic susceptibility at low temperatures there appears a cusp, characteristic for a spin-glass state (see e.g. References [26–28]) for $Hg_{1-x}Mn_xTe$, $Cd_{1-x}Mn_xTe$, and $Cd_{1-x}Mn_xSe$ with $x \geq 0.3$).

In III-V DMS the commonly observed high carrier concentration leads to high Curie temperatures. In thin films of $In_{1-x}Mn_xAs$ with carrier concentrations not exceeding $10^{19}$ cm$^{-3}$ antiferromagnetic Curie temperatures several times higher than in IV-VI DMS with comparable $x$ were observed.[29,30]

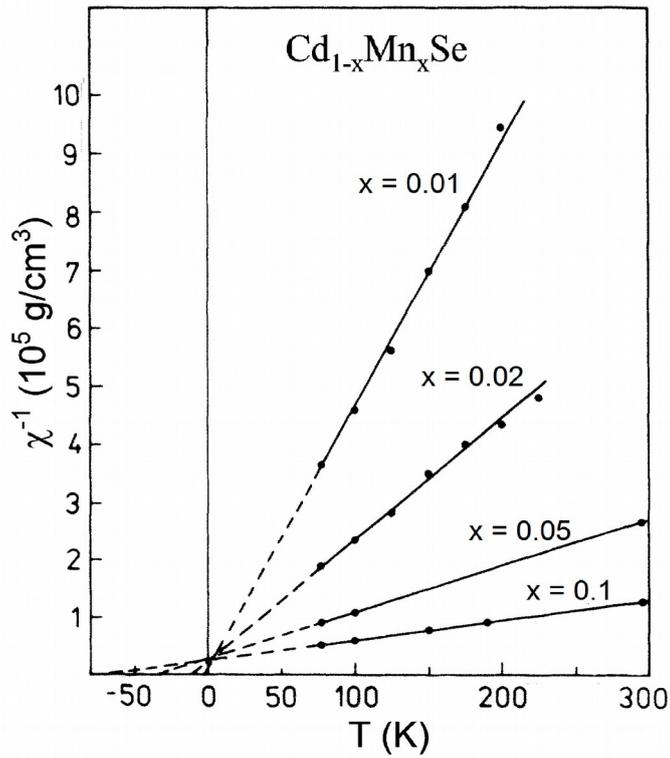

Figure 1. Inverse magnetic susceptibility vs temperature for $Cd_{1-x}Mn_xSe$. The lines are fits to the Curie-Weiss law. Adapted with permission.[24]

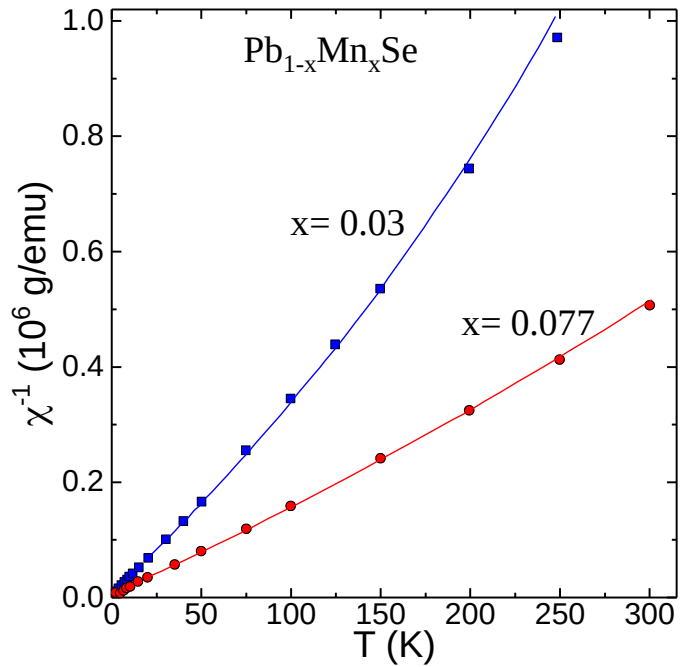

Figure 2. Inverse magnetic susceptibility vs temperature for $Pb_{1-x}Mn_xSe$. The lines are fits to the Curie-Weiss law. Adapted with permission.[25]

Magnetization of a system of magnetic ions can be expressed as a function of temperature and magnetic field:[31]

$$M = M_S + M_P + \chi_0 H, \tag{6}$$

where

$$M_S = N_0\, g_M\, \mu_B\, S\, \bar{x_1} B_S(\zeta). \tag{7}$$

$B_S(\zeta)$ is a modified Brillouin function:[32]

$$B_S(\zeta) = \frac{2S+1}{2S} \coth\left(\frac{2S+1}{2S}\zeta\right) - \frac{1}{2S} \coth\left(\frac{\zeta}{2S}\right), \tag{8}$$

and

$$\zeta = \frac{S\, g_M\, \mu_B\, H}{k_B (T + T_0)}. \tag{9}$$

The magnetization is usually given in emu/g or emu/cm$^3$, depending on the $N_0$ units. $S = 5/2$ for transition-metal-ions with half-filled $d$-shell and $7/2$ for rare-earth-ions with half-filled $f$-shell, $g_M = 2$ for Mn, Eu, Gd (in general $g_M$ may be anisotropic and different from 2, e.g. for Co $g_\parallel = 2.27$, $g_\perp = 2.29$).[33] The term $M_S$ represents the isolated magnetic ions, $\bar{x_1}$ is the effective number of the isolated magnetic ions in cation sites. $T_0$ is a phenomenological parameter representing the average contribution of magnetic ions to the exchange interaction, corresponding to $\theta$ in the molecular field approximation. When $M_P$ is omitted $T_0$ represents mainly the pair exchange interactions. If $T_0 = 0$, $M_S$ represents a system of non-interacting isolated ions.

The pair magnetization, derived from Equation (2), is:

$$M_P = \frac{1}{2} M_0 \acute{x}_2 \frac{\sum_{s=0}^{S_{max}} \exp\left[\frac{J_P}{k_B T} s(s+1)\right] \sinh\left(\frac{2s+1}{2s}\xi\right) s B_s(\xi)}{\sum_{s=0}^{S_{max}} \exp\left[\frac{J_P}{k_B T} s(s+1)\right] \sinh\left(\frac{2s+1}{2s}\xi\right)}, \tag{10}$$

where

$$\xi = \frac{s g_M \mu_B H}{k_B T}. \tag{11}$$

$S_{max} = 2S$, $\bar{x_2}$ represents the effective number of magnetic ions in pairs, and $J_P$ represents the pair exchange. The equation for $M_P$ is essentially the same as that given by Bastard and Leviner.[34]

For $J_P/k_B T \gg 1$ the term $M_P$ yields a steplike behavior (see Figure 3). For $J_P/k_B T \ll 1$ $M_P$ reduces to the Briloouin function, not the modified Brilouin function, since $\xi$ in $M_P$ is different from $\zeta$ in $M_S$, $M_P$ was not derived in the molecular field approximation. Such magnetization steps were first observed by Shapira *et al.* in $Cd_{1-x}Mn_xSe$ and $Zn_{1-x}Mn_xSe$,[35] later also in other DMS by Isaacs *et al.*,[36] Foner *et al.*,[37] and others.

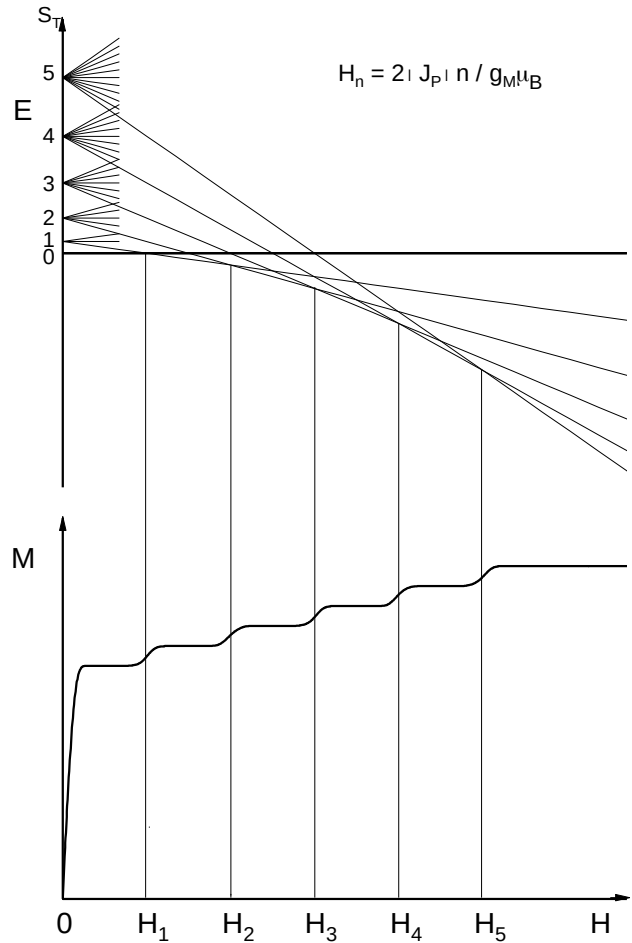

Figure 3. Schematic illustration of the formation of magnetization steps for Mn pairs in DMS. Upper part – energy level diagram (Equation (2)). Lower part – magnetization (Equation (6)).

A computer simulation of magnetization steps in $Pb_{1-x}Mn_xTe$ and $Cd_{1-x}Mn_xTe$ is shown in Figure 4. From Equation (2) we see, that the position of magnetization steps does not depend on $x$, but on $J_P$. It is important, because in mixed crystals the real magnetic ion content may be significantly different from the technological one. However, when $|J_P/k_B| < T$ we do not expect to see magnetization steps.

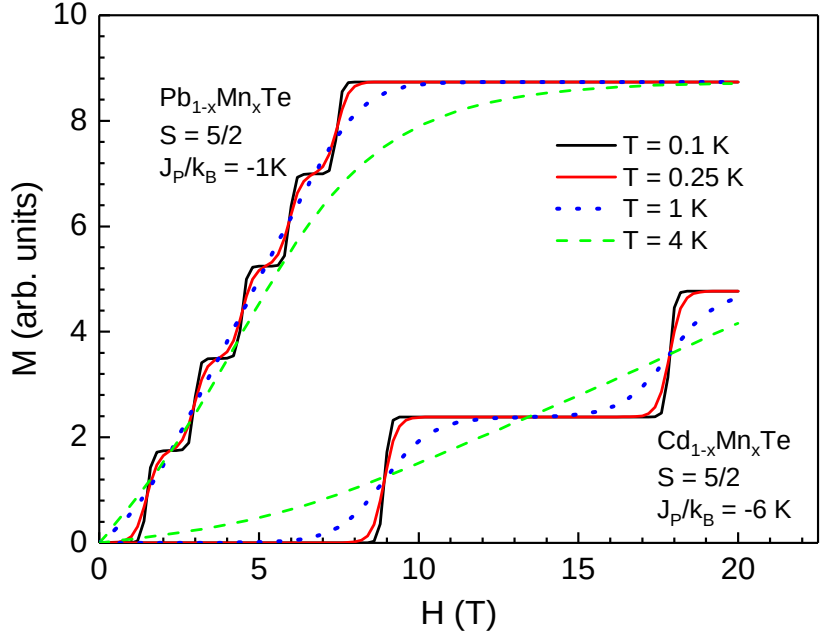

Figure 4. Computer simulation of magnetization of pairs only (Equation 10) for $Pb_{1-x}Mn_xTe$ and $Cd_{1-x}Mn_xTe$ with $x = 0.1$.

In Figures 5 and 6 we show high-field magnetization data for $Cd_{1-x}Mn_xTe$ and $Pb_{1-x}Mn_xTe$, respectively. In $Cd_{1-x}Mn_xTe$ the pair exchange parameter, $J_P/k_B$, determined from the magnetization steps agrees quite well with that determined from the susceptibility data. In $Pb_{1-x}Mn_xTe$ steps are not apparent in the data, but the magnetization versus magnetic field in the high-field region may be well described only when the pair-interaction function is included. From the fits to Equation (6) with all three components we obtained the pair exchange parameters in general agreement with the high-temperature susceptibility data.

From the results presented above we found that the exchange parameters in IV-VI DMS are about an order of magnitude smaller that in II-VI DMS. The only mechanism that can explain the observed difference is superexchange via an anion. In Reference [21] we have proposed a semi-empirical model of the superexchange interaction via an anion, based on the Anderson theory of the superexchange interaction[38] and the simple method of estimation of $J$ used by Spałek et al.[24] The model yields a very strong dependence of the superexchange interaction on the interatomic separation, $d$.

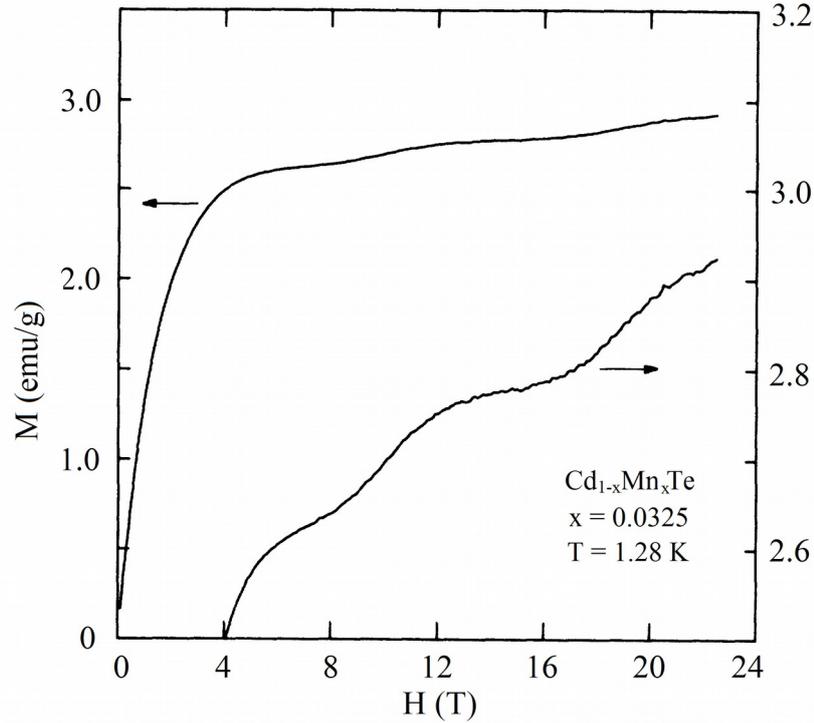

Figure 5. High-field magnetization of $Cd_{1-x}Mn_xTe$. The lower curve gives an expanded view of the first two steps seen in the upper curve. Adapted with permission.[36]

For transition metal ions:

$$J_{dpd} \sim \frac{r_p^2 r_d^6}{d^{16}}. \qquad (12)$$

For rare earth ions:

$$J_{fpf} \sim \frac{r_p^2 r_f^{10}}{d^{20}}. \qquad (13)$$

Here $r_p$ are tight-binding parameters of anion $p$ states, defined by Harrison and Straub,[39] $r_d$ and $r_f$ are parameters related to the cation $d$ and $f$ state radii,[18] $d$ is the cation-anion separation.

As we see, the superexchange via an anion should depend strongly on the cation-anion distance. For the same lattice constant, $a$, the cation-anion separation in the zinc-blende and wurtzite structure (II-VI and III-V compounds) equals $a\sqrt{3}/4$, while in the rocksalt structure (IV-VI compounds) it equals 0.5 $a$. The difference can lead to an order of magnitude or more difference in the superexchange parameter, $J$.

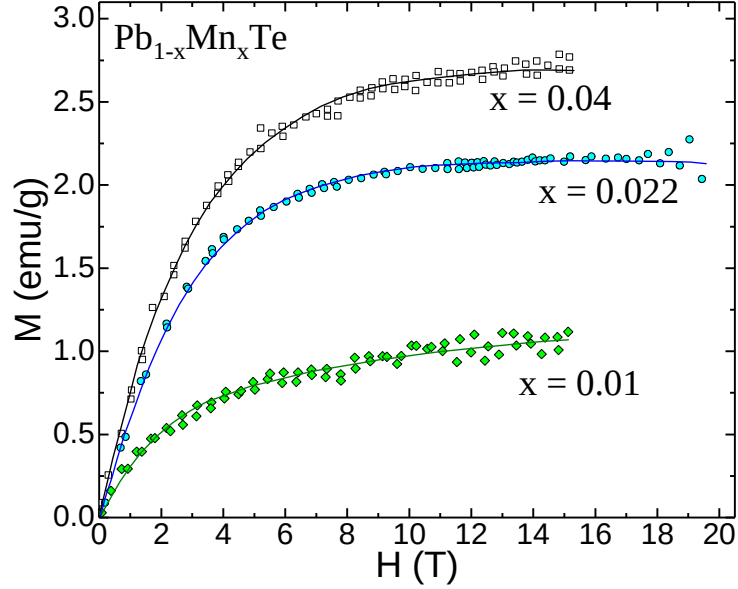

Figure 6. High-field magnetization of $Pb_{1-x}Mn_xTe$ at $T = 4.2$ K. The solid lines are fits to Equation (6). Adapted with permission.[31]

The estimated results for Mn-doped IV-VI chalcogenides are in good agreement with the experimental ones. In the rare-earth-doped systems, however, the estimated values for $J$ are much less than those actually observed. Kasuya, who encountered this problem in the europium chalcogenides suggested, that the superexchange interaction between the rare-earth ions includes an intra-atomic *f-d* interaction and interatomic *p-d* interaction.[40] The latter is much stronger than the *f-p* exchange, since the *f*-shell electrons are shielded and bound more closely to the nucleus than the *d*-shell electrons. Kasuya estimated that the combined *f-d* and *p-d* mechanism would result in a correct magnitude of the exchange interaction in rare-earth chalcogenides.

In many experiments the observed values of the exchange parameter in lead chalcogenides based DMS were in reasonably good agreement with those predicted by the model of *f-d* and *p-d* mechanism. Sometimes further corrections taking into account the mutual positions of the energies of *d*-levels – $E_d$, *f*-levels – $E_f$, and the top of the valence band – $E_v$ in the *p*-type materials were necessary, e.g. for $Pb_{1-x}Eu_xTe$.[41]

Magnetization steps in rare-earth-doped IV-VI DMS were shown in magnetization measurements at 20 mK.[42-48] The influence of clusters bigger than pairs was taken into account and discussed in detail.[49]

**Table 1.** Antiferromagnetic exchange parameters for ternary IV-VI DMS.

| Material | $-J_{NN}/k_B$ [K] | Experiment | References |
|---|---|---|---|
| $Pb_{1-x}Mn_xTe$ | 0.84 | magnetic susceptibility | [21] |
|  | 0.85 | magnetization | [31] |
| $Pb_{1-x}Mn_xSe$ | 1.67 | magnetic susceptibility | [21] |
|  | 0.98 | magnetization | [31] |
| $Pb_{1-x}Mn_xS$ | 0.54, 1.28[a)] | magnetization | [50], [21] |
| $Pb_{1-x}Ce_xTe$ | 0.20 | magnetization | [46] |
| $Pb_{1-x}Ce_xSe$ | 0.26, 0.71 | magnetization | [47], [51] |
| $Pb_{1-x}Ce_xS$ | 0.32 | magnetization | [48] |
| $Pb_{1-x}Eu_xTe$ | 0.32 | magnetic susceptibility | [41] |
|  | 0.26, 0.47 | magnetization | [42], [41] |
| $Pb_{1-x}Eu_xSe$ | 0.24 | magnetization | [43] |
| $Pb_{1-x}Eu_xS$ | 0.23 | magnetization | [44] |
| $Pb_{1-x}Gd_xTe$ | 0.36 | magnetic susceptibility | [21] |
|  | 0.45 | magnetization | [31] |
| $Pb_{1-x}U_xTe$ | 0.08 | magnetization | [52] |
| $Sn_{1-x}Eu_xTe$ | 0.31, 05 | magnetization | [45], [53] |
| $Sn_{1-x}Gd_xTe$ | 0.7 | magnetic susceptibility | [54] |
|  | 0.95 | magnetization | [54] |
| $Ge_{1-x}Eu_xTe$ | 0.61 | magnetization | [55] |

[a)] Calculated from $\theta$ in Ref. [50] according to Equation 5.

The values of the superexchange parameter in IV-VI DMS are an order of magnitude smaller than in II-VI and III-V DMS. An incomplete list of antiferromagnetic exchange parameters, $J_{NN}$, for ternary IV-VI DMS is given in Table 1. Only the representative data are listed. For detailed data see the References.

Another approach to account for long distance magnetic interactions is the Extended Nearest Neighbor Pair Approximation (ENNPA) model In this model one assumes long range magnetic interactions with an exchange integral of the form $J(R) = J_{NN}(R/R_0)^{-\kappa}$. Here $J_{NN}$ is a constant of the order of the exchange integral for nearest neighbors separated by a distance $R_0$ and $\kappa$ is an exponent characterizing the decrease of the exchange integral with increasing distance $R$ between two magnetic ions. The ENNPA was successfully applied in a number of materials to the description of the magnetization or magnetic specific-heat measurements (see, for example, Reference [56]).

### 3.2. Influence of free carriers on the magnetic system in IV-VI DMS

In II-VI and IV-VI DMS discussed above the concentration of free carriers did not exceed $10^{19}$ cm$^{-3}$. The results could be well interpreted by the superexchange mechanism via an anion.

Story *et al.* have shown that in *p*-type Pb$_{1-x-y}$Sn$_y$Mn$_x$Te as the carrier concentration increased the magnetic properties of the system changed from paramagnetic with a small antiferromagnetic interaction to clearly ferromagnetic.[12] The transition was rather sharp, occurring at the hole concentration, $p_{cr}$, about $3\cdot10^{20}$ cm$^{-3}$ and was reversible, i.e., when the hole concentration of the sample was reduced (by annealing in metal-rich vapor) the sample showed again the paramagnetic behavior. The authors explained the onset of the ferromagnetic behavior as due to an onset of the RKKY exchange interaction. The exchange parameter in the RKKY interaction has an oscillatory dependence on the carrier concentration and the distance between magnetic ions:[13]

$$J(R_{ij}) = \frac{m^* J_{sp-d}^2 a^6 k_f^4}{32\pi^3 \hbar^2} \left( \frac{\sin(2k_F R_{ij}) - 2k_F R_{ij} \cos(2k_F R_{ij})}{(2k_F R_{ij})^4} \right), \qquad (14)$$

where $m^*$ is the effective mass of carriers, $a$ is the lattice constant, $J_{sp-d}$ is free carrier – magnetic ion exchange parameter, $R_{ij}$ is the magnetic ion separation and $k_F$ is the Fermi vector ($k_F \sim p^{1/3}$).

In the two-band model of electronic structure Equation (14) would yield an exchange parameter gradually increasing with *p*, but not the threshold-like behavior, as observed in Reference [12]. Swagten *et al.* have explained the threshold in the exchange parameter at $p_{cr}$ by taking into account the more realistic band structure of SnTe, which should also be applied to Pb$_{1-x}$Sn$_x$Te with *x* over 0.7.[57] Schematic illustration of the band structure is shown in Figure 7. In SnTe there is an additional valley of the valence band, located on the Σ direction. The hole effective mass in the Σ-band is about 20 times greater than the effective mass in the L-band. When, for SnTe, *p* reaches the concentration $p_{cr} = 3\cdot10^{20}$ cm$^{-3}$, the Fermi level enters the Σ-band and heavy holes may take part in the exchange. From Equation (14) we see, that *J* is directly proportional to the effective mass, therefore at $p_{cr}$ *J* increases rapidly.

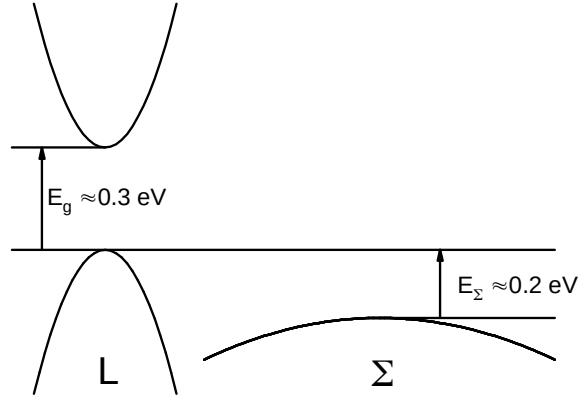

Figure 7. Schematic illustration of the band structure of SnTe and $Pb_{1-x}Sn_xTe$ for $x = 0.7$.

The ferromagnetic character of the exchange interaction in $Sn_{1-x}Mn_xTe$ was observed as early as in 1970 by Mathur et al.[10] The authors noticed the increase in the Curie temperature with increasing $x$, but did not relate it to the RKKY interaction. The systematic investigation of the carrier concentration dependence of the magnetic properties of was performed later by Eggenkamp et al.[58-60] In $Pb_{1-x-y}Sn_yMn_xTe$ an interesting effect of a carrier-induced ferromagnet-spin-glass transition was observed.[61] It was explained as due to the oscillatory character of the RKKY interaction. Story et al. presented a detailed analysis of RKKY interactions in many-valley IV-VI DMS including calculations of the exchange parameter and the Curie temperature.[62]

In $Sn_{1-x}Gd_{1-x}Te$, in our first experiments we did not observe any RKKY-type interaction, even in samples in which the hole concentration exceeded $10^{21}$ cm$^{-3}$.[63, 54] After changing the hole concentration in some samples (by annealing in Sn or Te vapor) we observed a strong enhancement in the Curie-Weiss temperature, $\theta$.[64,65] The enhancement in $\theta$ suggested an increase in the exchange parameter, $J$. That was later confirmed in the high-field magnetization measurements.[66] However, the influence of the hole concentration on the magnetic properties of $Sn_{1-x}Gd_{1-x}Te$ was qualitatively different from that described in $Pb_{1-x-y}Sn_yMn_xTe$. In $Pb_{1-x-y}Sn_yMn_xTe$ a change from antiferromagnetic to *ferromagnetic* interaction with increasing carrier concentration was observed. The effect of the hole concentration on the magnetic properties was *threshold-like*. In $Sn_{1-x}Gd_{1-x}Te$ we observed an enhancement in the *antiferromagnetic* interaction and the effect was *resonant*, observed only in samples with $0.02 < x < 0.05$ and hole concentration $2.5 \cdot 10^{20}$ cm$^{-3} < p < 4 \cdot 10^{20}$ cm$^{-3}$ (see Figure 8).

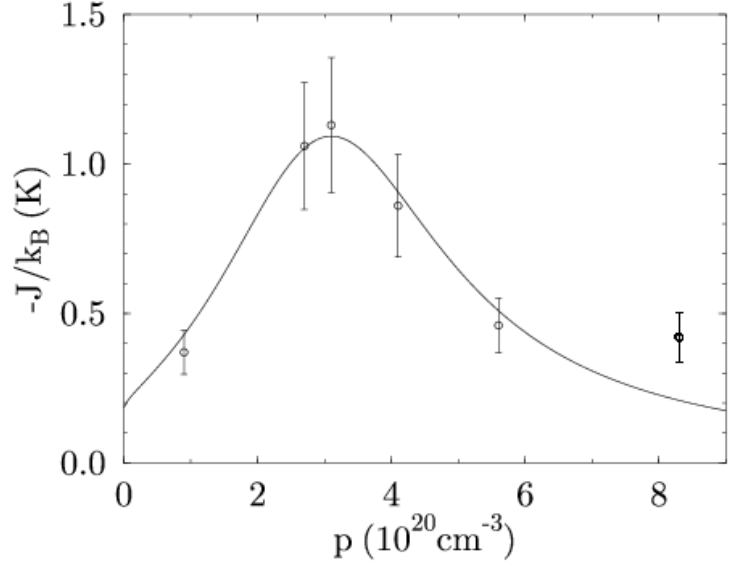

Figure 8. Exchange parameter vs hole concentration in $Sn_{1-x}Gd_xTe$. The solid line is the fit to the Lorentzian-type expression $-J/k_B = [C/(p^{2/3} - p_0^{2/3}) + \Gamma^2]$, based on a formula in which the $J_{ff} \sim (E_f - E_d)^{-2}$.[65] Here $C = 1.01$, $p_0 = 3.1$, $\Gamma = 0.96$. Reproduced with permission.[66]

The magnetic properties of rare-earth ions depend mostly on their *f*-shell electrons, which are shielded and bound closely to the nucleus. The occupied 4*f* levels are located several eV below the top of the valence band. Since the exchange interaction in rare-earth-doped DMS is a combined mechanism including an intra-atomic *f-d* interaction and an interatomic *d-p* interaction or an interatomic *d-d* interaction via free carriers, the position of the 5*d* level of the rare-earth ion in DMS becomes very important.

In lead chalcogenides based DMS the 5*d* level of the rare-earth ion is located in the conduction band and is not occupied. In those compounds the carrier concentration is below $5 \cdot 10^{19}$ cm$^{-3}$ and the Fermi level is located above the Σ valence band. The RKKY interaction is negligible and neither the threshold-like nor the resonant enhancement of the exchange interaction is observed.

In some $Sn_{1-x}Gd_xTe$ samples the 5*d* level of the Gd ion can be located in the valence band: (a) close to the top of the Σ-band, (b) close to the Fermi level, and be either occupied or not occupied. If both conditions (a) and (b) are satisfied, we observe a resonant RKKY-type exchange interaction via free holes with high effective mass. This case is schematically shown in Fig. 9. If the concentration of Gd is too high, the Gd 5*d* level is probably located above the top of the Σ-band and no enhancement due to the high effective mass is observed. Because of the complex character of the mechanism the exchange may be either ferromagnetic or antiferromagnetic. In our $Sn_{1-x}Gd_xTe$ crystals it was antiferromagnetic.

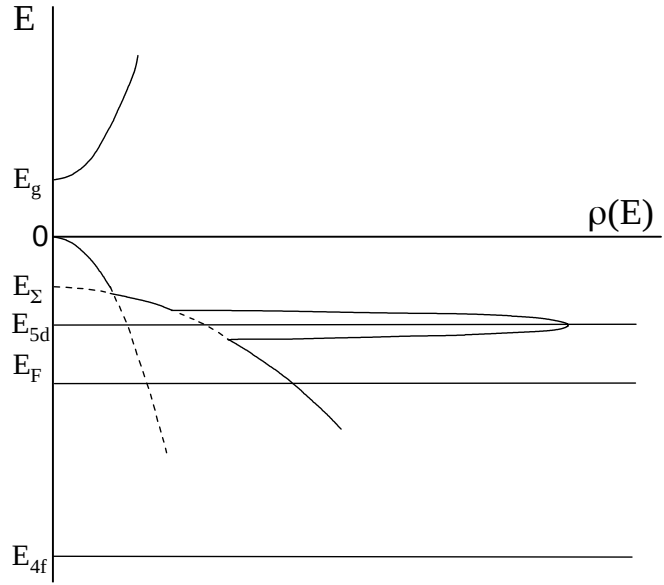

Figure 9. Schematic illustration of the band structure of $Sn_{1-x}Gd_xTe$, $\rho$ – density of states, $E_{4f}$ – Gd 4$f$ level, $E_F$ – Fermi level, $E_{5d}$ – Gd 5$d$ level, $E_\Sigma$ – top of the $\Sigma$-band, $E_g$ – energy gap. Adapted with permission.[64]

Let us summarize the results of magnetic susceptibility and magnetization measurements. We started with a simple model of a cation-anion-cation superexchange.[21] Then we had to add the intra-atomic *f-d* interaction and consider the mutual position of the rare-earth unoccupied 5$d$ level and the top of the valence band.[41] Taking into account the influence of free carriers we had to add the RKKY interaction.[12] Next we had to include the two valence band model of the energy band structure[57] and a presence of a rare-earth 5$d$ energy band resonant with the valence band.[64–66] As we will see in the next Section, this is not the end of additions necessary to understand the magnetic properties of IV-VI DMS.

## 4. Specific heat

A more complete information about exchange interactions may be obtained by measuring the magnetic specific heat. The results of magnetic specific heat measurements are complementary to the results of magnetic susceptibility and magnetization measurements. In the molecular field approximation, when a magnetic phase transition occurs at the critical temperature, $T_C$, there is a discontinuity at $T = T_C$ in the zero-field magnetic susceptibility and in the specific heat – the second derivatives of the free energy, with respect to magnetic field and temperature, respectively.[67] In experiments this effect is observed as a cusp in the temperature dependence of the susceptibility and the specific heat.

In the previous Sections we mentioned the phase transition from the paramagnetic phase to the spin-glass phase in II-VI DMS, due to the relatively large magnetic ion content,[26–28] and the transition from the paramagnetic to the ferromagnetic phase in IV-VI DMS due to the large carrier concentration.[12,57,58] A very good agreement between the paramagnet-ferromagnet transition observed by de Jonge *et al.* in the magnetic and the specific heat measurements of $Sn_{1-x}Mn_xTe$ is shown in Figure 10.[68] In DMS the transition temperature, $T_C$, usually matches the Curie-Weiss temperature.

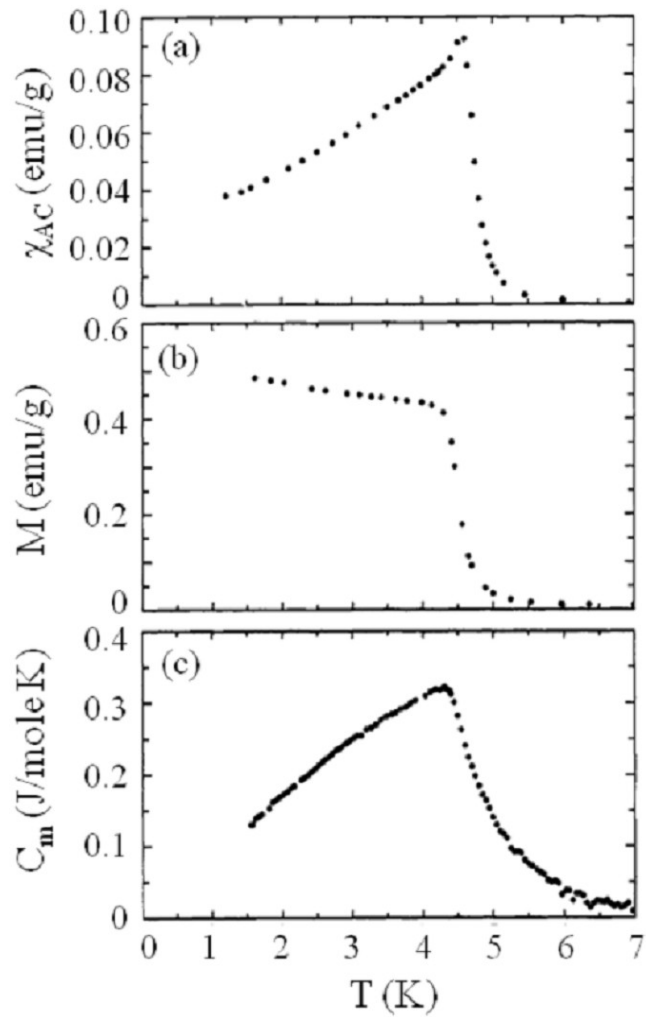

Figure 10. (a) AC susceptibility, (b) magnetization for $H_{DC}$ = 5 Gs, (c) magnetic specific heat of $Sn_{0.97}Mn_{0.03}Te$, $p = (7–10) \cdot 10^{20}$ cm$^{-3}$. Adapted with permission.[68]

In DMS based on lead chalcogenides, in our measurements of the magnetic susceptibility vs temperature above 2 K we did not observe the cusp characteristic for the transition to the spin-glass state. Escorne *et al.* observed a cusp in the magnetic susceptibility of $Pb_{1-x}Mn_xTe$ below 1 K.[69] In order to develop a more complete model of the exchange interaction among magnetic ions and to obtain parameters for this interaction we made complementary measurements of the magnetic specific heat of three DMS: $Pb_{1-x}Mn_xTe$, $Pb_{1-x}Eu_xTe$, and $Pb_{1-x}Gd_xTe$.[70-72]

In the analysis of the total specific heat it is not easy to separate the magnetic specific heat and the specific heat of the host lattice. The lattice specific heat is proportional to $T^3$ and at temperatures above 10 K significantly exceeds the magnetic contribution to the specific heat. Moreover, the specific heat of PbTe has an anomaly below 5 K and cannot be fitted with the standard expression $C = \gamma T + \alpha T^3$, where $\gamma T$ and $\alpha T^3$ are the electronic and lattice contributions, respectively.[73] The approach to this problem and the experimental details may be found, e.g. in Reference [70]. Below 5 K the lattice specific heat was about half of the total specific heat and in the interesting region, below 2 K, the specific heat of PbTe was more than 3 orders of magnitude smaller than that of PbTe based DMS.

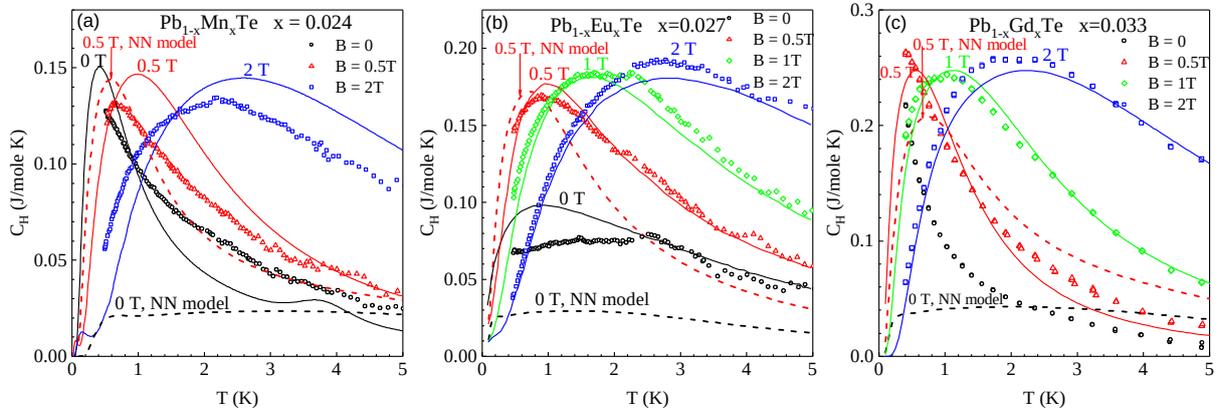

Figure 11. Magnetic specific heat of $Pb_{1-x}Mn_xTe$, $Pb_{1-x}Eu_xTe$, and $Pb_{1-x}Gd_xTe$. Markers – experimental data, dashed lines – NN exchange model (for B = 0 and B = 0.5T), solid lines - (a) Mn-carriers interaction model, (b) ground state splitting of $Eu^{2+}$ ions and $4f^7 \leftrightarrow 4f^65d^1$ transition, (c) 70% Gd ions are $Gd^{2+}$ in the $4f^75d^1$ configuration (see text). Adapted with permission.[70-72]

In Figure 11 we show the magnetic specific heat data for $Pb_{1-x}Mn_xTe$, $Pb_{1-x}Eu_xTe$, and $Pb_{1-x}Gd_xTe$ with similar magnetic ion content of about 0.03. The common and most interesting feature is the pronounced maximum around or below 2 K, in the absence of the magnetic field. In all materials the maximum is several times higher than that predicted by the model of superexchange between the nearest neighbors (NN) only, although the model described successfully the magnetic susceptibility and high field magnetization in all investigated materials, as mentioned above. For comparison we show the theoretical predictions of that model for $B$ = 0 and 0.5 T (broken lines in Figure 11), with the exchange parameters derived from the magnetic and magnetization data.[21,41] Below 0.5 T the maximum in $Pb_{1-x}Mn_xTe$ and $Pb_{1-x}Gd_xTe$ was below our experimental region (0.4 K). In higher fields the peak shifted towards higher temperatures.

We did not find one model describing well the specific heat in all three diluted magnetic semiconductors. Magnetic specific heat in $Pb_{1-x}Mn_xTe$ may be described in a model that takes into account the interaction between the *total spin systems* of the manganese ions and free carriers.[70] The model for $Pb_{1-x}Eu_xTe$ takes into account the single Eu ions with the ground state split in a disordered crystal field potential and the virtual $4f^7 \leftrightarrow 4f^65d^1$ transition, where the $5d^1$ level lies high in the conduction band.[71] In $Pb_{1-x}Gd_xTe$ the $5d^1$ level lies close to the bottom of the conduction band and as the Gd concentration increases we may observe the co-existence of $Gd^{3+}$ ions with the $4f^7$ configuration and the $Gd^{2+}$ ions with the $4f^75d^1$ configuration. The model taking into account the hybridization between the gadolinium $5d^1$ level and crystal band states describes well the magnetic specific heat for $B > 0$, where the peaks appear at magnetic fields *lower* than the NN model predicted, but not at $B$ = 0.[72] Detailed descriptions of the three models are in References [70-72].

Now we see that to understand exchange interactions in IV-VI DMS we must add to the interactions described in the previous Sections interactions of the total magnetic and electronic systems, in $Pb_{1-x}Mn_xTe$, as well as virtual and non-virtual transitions of electrons between the $4f^7$ and $5d^1$ levels in $Pb_{1-x}Eu_xTe$ and $Pb_{1-x}Gd_xTe$, respectively.

Another experimental method providing information about magnetic properties of semiconductors are measurements of neutron diffraction. We will not discuss it here. Vennix *et al*. performed neutron diffraction measurements of $Sn_{0.97}Mn_{0.03}Te$ with $p$ = (7–11) $\cdot 10^{20}$ cm$^{-3}$.[74] The results were in good agreement with results of earlier measurements of magnetization, susceptibility, and magnetic specific heat, where the carrier-induced paramagnet-ferromagnet phase transition was observed.[61]

## 5. Influence of magnetic ions on the electronic system in DMS

DMS are attractive because of the mutual influence of the electronic and magnetic subsystems. In Sections 3 and 4 we wrote about interactions among magnetic ions due to the influence of free carriers on the magnetic system in IV-VI DMS. We must not forget the influence of magnetic ions on the electronic system. When the semiconductor contains localized magnetic moments, its band structure will be modified by the exchange interaction of these moments with band electrons (for transition metals also called *sp-d* interaction). The electron-ion exchange may lead to a reduction or enhancement in the spin splitting of the Landau levels, reverse in the sign of the spin splitting and mixing of the Landau level sequence.[3]

The exceptionally large spin splitting of the band edges and of Landau, impurity, or electronic levels has a significant influence on the optical and electronic properties of DMS and leads to new, unique effects such as giant Faraday rotation,[75, 76] magnetically induced semimetal-to-semiconductor transition,[77] or negative magnetoresistance, resulting even in magnetic-field-induced metal-to-insulator transition,[78] At the early stage of research these effects were observed mostly in II-VI DMS and investigated by magnetooptical and magnetotransport experiments. Good review articles can be found in References [4, 79, 80]

IV-VI DMS are mostly narrow gap semiconductors with carrier concentrations of an order of $10^{18}$ cm$^{-3}$, or above. They can be either *n*- or *p*-type and usually are semimetallic. Magnetooptical experiments in these materials were carried out in the mid- and far-infrared. Early measurements of the cyclotron resonance in $Pb_{1-x}Mn_xTe$ bulk crystals[81] and thin films[82] have shown the influence of magnetic ions on the carrier effective masses and effective g-factors. Pascher *et* al. measured the coherent anti-Stokes Raman scattering (CARS) in $Pb_{1-x}Mn_xTe$ thin films and presented a detailed analysis of the band structure parameters in this material.[83] Geist *et al.* obtained similar results for $Pb_{1-x}Mn_xSe$.[84]

With development of the MBE technique it became possible to obtain good quality thin films of IV-VI DMS, suitable for magnetooptical experiments. Magnetoptical properties of IV-VI DMS based on lead chalcogenides and a good analysis of *sp-d* exchange interaction are given in the review by Bauer *et al*.[85] Other information and reviews may be found in References [9, 86-88]. Photoluminescence in IV-VI thin films from a point of view of applications for light emitting diodes and semiconductor lasers was investigated by Partin *et al*.[89]

Magnetotransport experiments included measurements of the Shubnikov-de-Haas (ShdH) effect,[82] the magnetoresistance and the Hall effect. A strong anomalous Hall effect (AHE) was observed in PbSnMnTe crystals with a high Mn content.[90] Analysis of the diluted magnetism in the magnetotransport properties of bulk IV-VI DMS is not as conclusive as in the case of II-VI DMS due to the degenerate nature of carrier conduction.[2, 9, 91]

In the present review we will not describe magnetooptical and magnetotransport experiments. Some important results of magnetotransport and magnetooptical experiments in thin films and heterostructures of IV-VI DMS will be reported in Section 7.

## 6. Spinodal decomposition and clusters

### 6.1. Ternary IV-VI DMS

In the previous Sections, for the description of magnetic properties of DMS we used models assuming a random distribution of magnetic ions in the cation sublattice. In such case the virtual crystal approximation and the molecular field approximation can be successfully applied for phenomenological description and theoretical analysis of the compound properties. However, even in ternary compounds such model happens to be unrealistic, and becomes less realistic with an increase of the magnetic ion content or in more complex materials such as quaternary compounds.

Below we will show magnetic properties of bulk IV-VI DMS materials in which deviations from random distribution of magnetic ions were observed. The non-random distribution may take a form of spinodal decomposition, where magnetic ions form clusters of higher content than the average in the matrix, but still preserve the same structure as the non-magnetic host material. There may be also clusters with structure different from that of the host material. A comprehensive review of spinodal nanodecomposition in semiconductors doped with transition metals may be found in Reference [92].

In our measurements of magnetic properties of $Pb_{1-x}Mn_xTe$ and $Pb_{1-x}Eu_xTe$ the analysis of magnetization in high magnetic field indicated some deviations from random distribution of magnetic ions, but in structural measurements no inhomogeneities were observed. In $Sn_{1-x}Mn_xTe$ phase transitions to spin-glass state or ferromagnetic phase were observed, but no inhomogeneities in the crystal structure were reported. $Sn_{1-x}Eu_xTe$ crystals with $x < 0.014$ were homogeneous, but for $x \geq 0.014$ a cusp in the susceptibility vs temperature plot appeared at about 10 K.[93] It resembled a paramagnet – spin-glass transition observed before in other DMS, but turned out to be due to the formation of antiferromagnetic EuTe inclusions. Another feature characteristic of antiferromagnetic clusters was a knee in the high field magnetization data at about 7 T, with approximately linear behavior below and above the knee.[94]

Above we discussed magnetic properties of several IV-VI DMS based on lead and tin chalcogenides. Now we will include a group based on germanium telluride. Electronic, magnetic, and optical properties of GeTe were described by Lewis.[95] GeTe like SnTe is a *p*-type semiconductor with a high carrier concentration – typically of an order of $10^{20}$ - $10^{21}$ cm$^{-3}$. Unlike PbTe and SnTe (at $T > 80$ K), GeTe at temperatures below 700 K has not a rocksalt but a rhombohedral structure. It is a distorted cubic structure with a corner angle of about 88˚. The rhombohedral deformation of the NaCl structure in IV-VI semiconductors occurs at temperatures roughly inversely proportional to the cation mass. The transition from the cubic to the rhombohedral phase for GeTe is at about 700 K,[96] for SnTe is below 100 K (down to zero for high carrier concentrations),[97, 98] and PbTe is always cubic.

In 1974 Cohrane *et al.* measured magnetic properties of bulk $Ge_{1-x}Mn_xTe$ with $0 < x < 0.5$. The compounds were ferromagnetic with Curie temperatures as high as 150 K for $x = 0.5$.[99] Results for $x < 0.15$ could be well explained by the RKKY interaction, as for $Sn_{1-x}Mn_xTe$. For higher Mn contents there were anomalies in the temperature dependence of magnetization. The authors mentioned possible inhomogeneities in Mn distribution, but suggested other explanations of the effect as well. We studied $Ge_{1-x}Eu_xTe$ bulk crystals with the chemical composition, *x*, changing from 0.008 to 0.025.[55] In the samples with $x < 0.015$ we did not see clusters with the NaCl structure but spinodal decompositions with the structure of the host lattice. Some of these inhomogeneities had Eu-contents much higher than the average *x* value. The solubility of Eu in the GeTe lattice turned out to be below 0.01. The structural analysis of the samples with $x \geq 0.020$ indicated a presence of EuTe with NaCl structure. There we observed a cusp in the magnetic susceptibility vs temperature plot at about 11 K and a knee in the magnetization vs magnetic field plot at about 7 T like in $Sn_{1-x}Eu_xTe$. This we attributed to the formation of an antiferromagnetic EuTe phase. For concentrations less than 0.015 we did not observe a cusp or direct evidence of other Eu-rich inclusions. However, the analysis of the high-field magnetization in all samples indicated a non-random distribution of magnetically active Eu-ions, especially very few isolated ions.

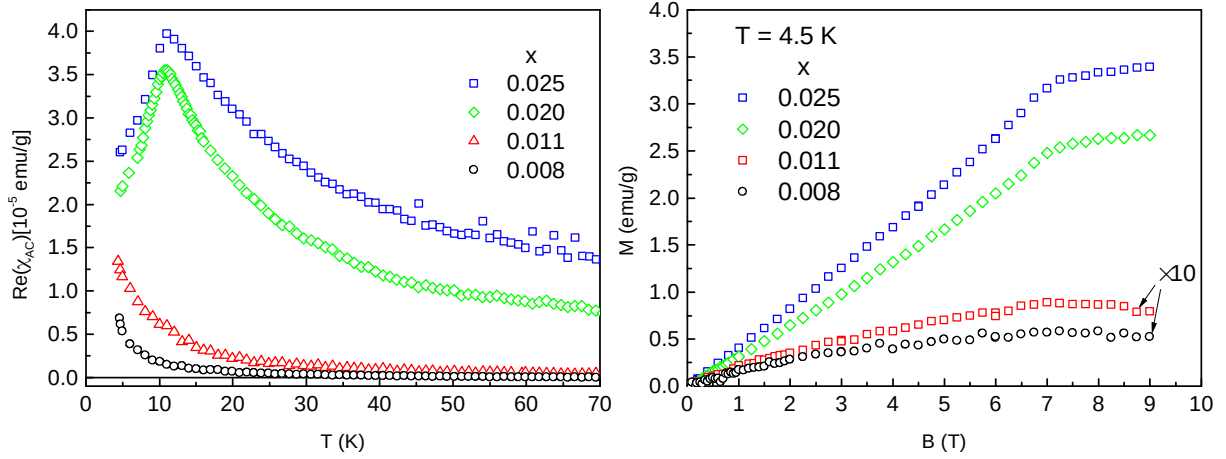

Figure 12. Magnetic properties of $Ge_{1-x}Eu_xTe$ (a) temperature dependence of ac magnetic susceptibility, (b) magnetic field dependence of magnetization. Adapted with permission.[55]

In Figure 12 we show typical results indicating presence of EuTe clusters in Eu-doped IV-VI DMS.

In the IV-VI DMS described above the most investigated transition metal ion was manganese and the most investigated rare-earth-ion was europium. Both are isoelectronic dopants in IV-VI semiconductors and have half-filled magnetic orbitals – $3d^5$ for Mn and $4d^7$ for Eu. In such configuration the orbital magnetic moment $L = 0$, and the total magnetic moment $J = S$, we may use $J$ and $S$ interchangeably (as we did). There is as well "after-manganese life" and one of the possibilities is chromium. Cr is also an isoelectronic dopant, but the magnetic orbital configuration of the $Cr^{2+}$ ion is $3d^4$. Now $L$ is nonzero and generally $J \neq S$. In semiconductors often the ligand crystal field influence leads to quenching of the orbital contribution to the magnetic moment and it is still correct to assume $J = S$. There is evidence of $Cr^{2+}$ with $J = S = 2$ in $Ge_{1-x}Cr_xTe$ thin films.[100] The solubility of chromium in IV-VI materials is less than 0.01 in bulk PbTe,[101] 0.012 in bulk SnTe,[102] and up to 0.06 in bulk GeTe.[103]

Mac *et al*. measured magnetization of $Pb_{1-x}Cr_xTe$ with $x$ up to 0.01.[101] They observed a Brillouin-like paramagnetic behavior and the analysis indicated a presence of $Cr^{2+}$ and $Cr^{3+}$ ions. The authors mentioned a possibility of clustering, but no direct evidence of either spinodal decomposition or other clusters. Magnetic properties of $Pb_{1-x}Cr_xTe$ with $x$ below 0.01 were recently studied by Gas *et al*.[104] The presence of $Cr^{2+}$ and $Cr^{3+}$ ions was observed, as well as the presence of $Cr_2Te_3$ and $Cr_5Te_8$ clusters. In $Sn_{1-x}Cr_xTe$ we observed Cr-ion rich regions, i.e. spinodal decomposition.[102] They were not related to any $Cr_{1-\delta}Te$ phases. These inhomogeneities led to formation of a spin-glass-like state at temperatures below 140 K. (We call it spin-glass-like to distinguish from the spin-glass state observed in homogeneous DMS discussed above; it is also called cluster-glass). The peaks corresponding to the transition were symmetrical, therefore we knew we did not observe paramagnet-ferromagnet transitions, then the peaks would be non-symmetrical, like in Figure 10 in the case of $Sn_{1-x}Mn_xTe$.[61] Analysis of the frequency dependence of the ac magnetic susceptibility let us to identify the observed state as spin-glass-like, not superparamagnetic.

Magnetic properties of $Ge_{1-x}Cr_xTe$ were more complicated.[103] No Cr-rich regions were found in samples with $x < 0.05$, spinodal decompositions were observed for $x \geq 0.05$. For samples with $x < 0.03$ the magnetic susceptibility and magnetization measurements indicated a presence of a spin-glass state, while for samples with $x > 0.03$ a ferromagnetic state was observed. The carrier concentration was of an order of $3 \cdot 10^{20}$ cm$^{-3}$. We found that the RKKY interaction was responsible for the observed magnetic ordering in samples with $x < 0.045$, and the magnetic properties of the samples with $x > 0.045$ were dominated by the presence of spinodal decompositions.

The results led to two conclusions:

1. Spinodal decomposition and clusters were least likely to form in IV-VI DMS with the rocksalt structure, like PbTe-based DMS and most likely to form in IV-VI DMS with the rhombohedral structure, like GeTe-based DMS, with SnTe-based DMS in-between.

2. Cr-doped IV-VI DMS were most likely to form spinodal decomposition. We observed that the magnetic properties of $Sn_{1-x}Cr_xTe$ resembled more those of $Ge_{1-x}Cr_xTe$ than those of $Sn_{1-x}Mn_xTe$.

### 6.2. Quaternary IV-VI DMS

Chromium is difficult to be introduced into the IV-VI compound lattice. We wanted to see whether an addition of Pb or Eu to $Sn_{1-x}Cr_xTe$ would improve chromium solubility. The result was quite opposite. There appeared clusters with mostly Cr and Te ions.

The magnetic properties of $Sn_{1-x-y}Pb_yCr_xTe$ and $Sn_{1-x-y}Eu_yCr_xTe$ crystals were similar. The magnetic susceptibility vs temperature for the $Sn_{1-x-y}Pb_yCr_xTe$ samples with $x \approx 0.01$ and $Sn_{1-x-y}Eu_yCr_xTe$ with $x < 0.07$ showed the presence of a well-defined peak at temperatures around 120 – 140 K.[105,106] This effect was identified as the appearance of a cluster-spin-glass freezing of the $Cr_2Te_3$ grains. The possible presence of such grains was confirmed by the diffraction patterns. In the ternary $Sn_{1-x}Cr_xTe$ with $x \leq 0.012$ we observed a similar transition, but there we did not see any evidence of clusters with the phase other than the matrix – only the spinodal decomposition. The temperature dependence of the magnetic susceptibility for the samples with $x \geq 0.07$, both in $Sn_{1-x-y}Pb_yCr_xTe$ and $Sn_{1-x-y}Eu_yCr_xTe$, showed a paramagnet-ferromagnet transition located at temperatures close to 300 K. The ferromagnetic alignment of Cr-ions in these samples was related to the presence of $Cr_5Te_8$ clusters. The addition of either Pb or Eu did not improve the solubility of Cr in SnTe.

For all samples with Eu ions we have seen a paramagnetic contribution to the total magnetic susceptibility at temperatures below 50 K. This paramagnetic contribution to the magnetic susceptibility seems to be related to the Eu and Cr ions randomly distributed in the SnTe lattice. It was not observed in samples without europium.[106]

Next, we experimented with adding Eu or Pb to $Ge_{1-x}Cr_xTe$. The $Ge_{1-x-y}Cr_xEu_yTe$ samples with $0.015 \leq x \leq 0.057$ and $0.03 \leq y \leq 0.042$ crystallized in the rhombohedrally distorted GeTe structure and were single phased. Spinodal decompositions into regions with low and high Cr contents were observed. In the magnetic susceptibility vs temperature dependence we observed a typical paramagnet – ferromagnet transition at the temperature around 50 K in all samples, except one with the lowest Eu content, where the peak was symmetrical, typical of the paramagnet – spin glass transition.[107] The transition temperatures were very close to the transition temperatures in $Ge_{1-x}Cr_xTe$ and did not depend on the Eu-content.

All the $Ge_{1-x-y}Pb_yCr_xTe$ materials with $0.017 < x < 0.043$ and $0.18 < y < 0.22$ were composite compounds, there were PbTe-based nanoscale dots and $Cr_5Te_8$ clusters. In all samples not *one* but *two magnetic phase transitions* were found in the studied system, one between 130 K and 150 K and the other around 40 K. The high-temperature transition was a paramagnet to spin-glass freezing, while the low temperature transition changed its character with the Cr content. For $x \leq 0.035$ a paramagnet to spin-glass freezing was observed, while for $x > 0.035$ the transition was paramagnet – ferromagnet.[108] We believe that the low-temperature spin-glass state or ferromagnetic phase is related to the matrix, while the high-temperature spin-glass state is related to the $Cr_5Te_8$ clusters.

Here we may recall the two phase transitions depending on carrier concentration in $Pb_{1-x-y}Sn_yMn_xTe$: for $p$ below $10^{21}$ cm$^{-3}$ a paramagnet-ferromagnet and for $p$ above $10^{21}$ cm$^{-3}$ a ferromagnet-spin-glass transition.[12, 61] This effect was not related to inhomogeneities but to the oscillatory character of the RKKY interaction.

We also experimented with adding Pb, Sn, or Eu to $Ge_{1-x}Mn_xTe$.[109-111] Magnetic properties of such quaternary compounds were similar to those described above for $Ge_{1-x}Cr_xTe$ co-doped with Pb and Eu.

We noted that no antiferromagnetic EuTe clusters either in $Sn_{1-x-y}Eu_yCr_xTe$ or in $Ge_{1-x-y}Cr_xEu_yTe$ were observed, although such clusters appeared both in $Sn_{1-x}Eu_xTe$ and in $Ge_{1-x}Eu_xTe$ with the Eu content lower than that in the quaternary IV-VI DMS, as shown above.

Summarizing, as a result of spinodal decomposition and/or clusters of different structure in the IV-VI DMS we observed an antiferromagnetic phase ($Sn_{1-x}Eu_xTe$, $Ge_{1-x}Eu_xTe$), a spin-glass-like state ($Sn_{1-x}Cr_xTe$, $Ge_{1-x-y}Sn_yMn_xTe$), a spin-glass state or a ferromagnetic phase, depending on the magnetic ion content ($Sn_{1-x-y}Pb_yCr_xTe$, $Sn_{1-x-y}Cr_xEu_yTe$, $Ge_{1-x}Cr_xTe$, $Ge_{1-x-y}Cr_xEu_yTe$), two spin-glass states in two different temperature ranges in the same sample ($Ge_{1-x-y}Pb_yMn_xTe$), a spin-glass in one temperature range and a ferromagnetic phase in another temperature range in the same sample ($Ge_{1-x-y}Pb_yCr_xTe$, $Ge_{1-x-y}Eu_yMn_xTe$). Some examples are shown in Figures 12 and 13.

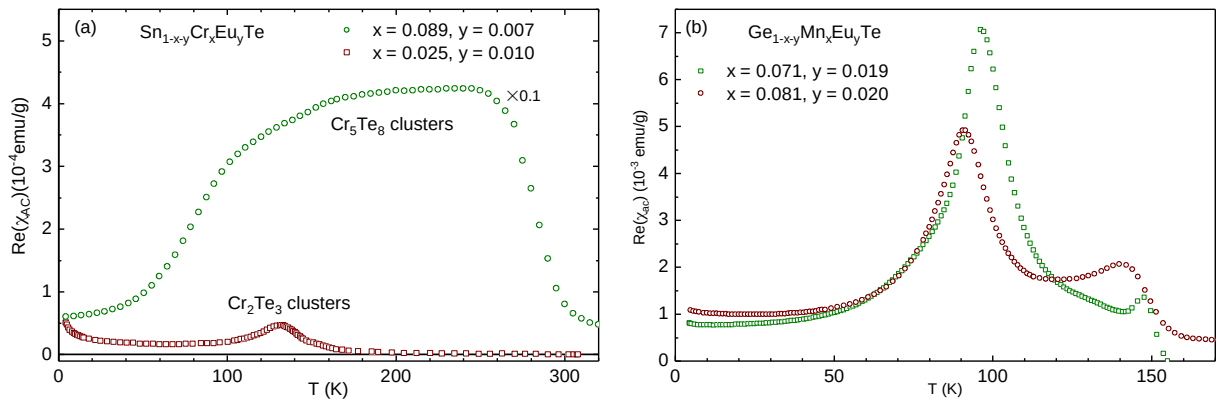

Figure 13. Temperature dependence of ac magnetic susceptibility for two quaternary IV-VI DMS with different magnetic phases, (a) spin-glass or ferromagnetic, (b) spin-glass and ferromagnetic. Adapted with permission.[106,111]

## 7. Thin films and low-dimensional systems

### 7.1. Thin films of IV-VI DMS based on lead and tin chalcogenides

In the previous Sections we described magnetic properties of bulk IV-VI DMS. As was mentioned above, a certain drawback in the research was the relatively low solubility of transition metals and rare-earth-ions in bulk IV-VI semiconductors, comparing to II-VI semiconductors. Similar problem was encountered in the research on III-V DMS. A breakthrough was made by using the molecular beam epitaxy (MBE) technique.[14,112] A successful epitaxial growth of III-V DMS was reported thereafter in many papers.

MBE growth technique was also successfully applied for IV-VI DMS. Braunstein *et al.* investigated magnetic properties of $Pb_{1-x}Eu_xTe$ with $x$ up to 0.31 and results were similar to those described in Section 3.1.[113] Krenn *et al.* studied magnetic properties of $Pb_{1-x}Eu_xTe$ in the concentration range $0 < x \leq 1$ by means of magnetooptical and magnetic measurements.[114] Partin *et al.* used the MBE growth technique to obtain a wide series of lead-rare-earth chalcogenides: doped with Yb,[115] Ho. Er, Dy,[116] and Eu.[117] These materials were investigated not with respect to their magnetic properties but to the possibility of tuning their narrow energy gap. They could be useful for fabricating heterojunction diode lasers in the spectral range from 1 to 5 μm.

Nadolny *et al*. prepared $Sn_{1-x}Mn_xTe$ films with $x$ up to 0.04 by the MBE technique on $BaF_2$ (111) substrates with SnTe buffer layer.[118] An extensive study of magnetotransport in $Sn_{1-x}Mn_xTe$ ($0.00 \leq x \leq 0.08$) (111) epitaxial thin films grown by MBE on $BaF_2$ substrates was presented by Adhikari *et al.*[119] Magnetic properties of the $Sn_{1-x}Mn_xTe$ films were similar to those of the bulk materials. Ishida *et al.* prepared $Sn_{1-x}Eu_xTe$ and $Sn_{1-x}Cr_xTe$ films with $x$ up to 0.06 by the hot wall epitaxy and described the electrical and optical properties of the layers.[120]

## 7.2. Thin films of IV-VI DMS based on GeTe

With development of spintronics more attention was paid to the thin films of IV-VI DMS with high carrier concentrations. Most of them were based on GeTe. Fukuma *et al.* carried out preparations and investigations of GeTe based DMS in wide concentration range such as $Ge_{1-x}Mn_xTe$ with $x$ up to 0.96[121] and $Ge_{1-x}Cr_xTe$ with $x$ up to 0.33.[100] They used the cluster beam and rf sputtering techniques. By the rf sputtering on glass substrates Fukuma *et al.* prepared thin films of $Ge_{1-x}TM_xTe$ with $x$ about 0.15, where TM were transition metals from Ti to Ni.[122] The authors investigated magnetotransport and magnetic properties of the materials. X-ray diffraction patterns were obtained for all materials. Films of GeTe with Ti, Cr, Mn, and Co had the GeTe structure with no other phases. The diffraction patterns of V, Fe, and Ni doped GeTe films showed a presence of second phases of binary $TM_{1-\delta}Te$ alloys. Magnetic measurements indicated that Ti, V, Co, and Ni doped GeTe films were paramagnetic, while Cr, Mn, and Fe doped GeTe films were ferromagnetic with Curie temperatures 12, 47, and 100 K, respectively. The authors attributed the ferromagnetic ordering to the RKKY-type interactions, although in $Ge_{1-x}Fe_xTe$ there were clusters of $Fe_{1-\delta}Te$ alloys.

Asada *et al.* experimented with quaternary thin films $Ge_{1-x-y}Pb_yMn_xTe$ with $0.05 \leq x \leq 0.3$ and $y$ roughly from 0 to 1.[123] By changing Pb content they controlled the carrier concentration and by changing Mn content they adjusted the magnetic interactions. They reported the Curie-Weiss temperature of about 90 K for $x = 0.22$ and $y = 0.47$.

From the point of view of spintronics the most prospective materials seemed to be $Ge_{1-x}Mn_xTe$ and $Ge_{1-x}Cr_xTe$. Fukuma *et al.* investigated magnetic properties of $Ge_{1-x}Mn_xTe$ and $Ge_{1-x}Cr_xTe$ thin films deposited by the MBE technique on $BaF_2$ substrates. For $Ge_{1-x}Cr_xTe$ with $x \approx 0.06$ they observed the Curie temperature of 180 K.[124] For $Ge_{1-x}Mn_xTe$ with $x \approx 0.08$ they observed the Curie temperature of 190 K.[125] The Curie temperatures are comparable to those reported for III-V DMS.

GeTe has a rhombohedral structure, with a corner angle of about 88°, below 700 K. However, with adding Mn the corner angle increases continuously and reaches 90° at $x$ about 0.18. From 0.18 to 0.60 the phase is the high-temperature cubic NaCl. The lattice parameter also varies smoothly from 5.98 Å in pure GeTe to 5.88 Å. The structure is shown in Figure 14. Beyond $x \sim 0.9$ the alloys crystallize into the hexagonal NiAs structure of MnTe, while the region between 0.6 and 0.9 is two-phase. These features were described by Cohrane *et al.* for bulk $Ge_{1-x}Mn_xTe$.[99] Fukuma *et al.* in their $Ge_{1-x}Mn_xTe$ thin films on $BaF_2$ substrates also observed the change from the rhombohedral to the NaCl structure at $x \sim 0.2$, but their films retained the cubic structure up to $x \sim 0.96$.[121]

Chen *et al.* prepared $Ge_{1-x}Mn_xTe$ films with $0.26 \leq x \leq 0.98$ by the MBE technique on $BaF_2$ (111) substrates.[126] The films crystallized in the NaCl phase with (111) orientation when $x > 0.5$ and tended to be amorphous for $x < 0.5$. All films were ferromagnetic and the Curie temperature was about 90 K for $x = 0.25$, increased with $x$ reaching the maximum of 110 K at $x = 0.67$, and then decreased down to about 80 K for $x = 0.98$. The authors attributed this behavior to the competition between the ferromagnetic RKKY-type interaction and antiferromagnetic superexchange among the Mn ions. In subsequent studies Chen *et el.* attained the Curie temperature of 180 K for $x = 0.55$.[127]

Lechner *et al.* have shown that the structural and magnetic properties of $Ge_{1-x}Mn_xTe$ depend on the epitaxial growth conditions.[128] Even small changes in the growth temperature lead to the formation of secondary phases of antiferromagnetic hexagonal MnTe and rhombohedral $Ge_{1-x}Mn_xTe$ with reduced $x$. The phase separation lead to a reduced content of the cubic ferromagnetic $Ge_{1-x}Mn_xTe$ phase in the layers, and the coexistence of the ferromagnetic and antiferromagnetic phases decreased the magnetization. Later Hassan *et al.* carried out a systematic growth study of $Ge_{1-x}Mn_xTe$ in which Mn content as well as growth conditions were changed.[129] They concluded that only within a narrow window of growth conditions single phase ternary $Ge_{1-x}Mn_xTe$ layers could be obtained. At lower or higher growth temperatures, precipitation of secondary phases or even phase separation occurred. Under optimized conditions, high quality $Ge_{1-x}Mn_xTe$ layers were obtained with Curie-Weiss temperatures as high as 200 K for Mn content $x = 0.46$.

Knoff *et al.* intentionally produced polycrystalline $Ge_{0.915}Mn_{0.085}Te$ ferromagnetic semiconductor microstructures embedded in an amorphous, paramagnetic (Ge,Mn)Te matrix. [130] The microstructures were produced by pulsed laser and electron beam induced local re-crystallization of amorphous layers. The re-crystallized regions were ferromagnetic (Ge,Mn)Te thin disks of submicron dimensions, with the Curie temperature of 70 K.

In Figure 14 we recall the crystalline structure of $Ge_{1-x}Mn_xTe$. For $x < 0.18$ it is rhombohedral, while for $0.18 \leq x \leq 0.60$ it is cubic, NaCl structure. The rhombohedral GeTe and $Ge_{1-x}Mn_xTe$ are *ferroelectric* due to the relative displacement of the cation and anion sublattices along the <111> direction and the lack of inversion symmetry. However, due to the RKKY interactions among magnetic ions, described above, $Ge_{1-x}Mn_xTe$ is also *ferromagnetic*. The combination of these two properties in a lattice with only two atoms in the primitive unit cell makes $Ge_{1-x}Mn_xTe$ the simplest *multifferroic* material.

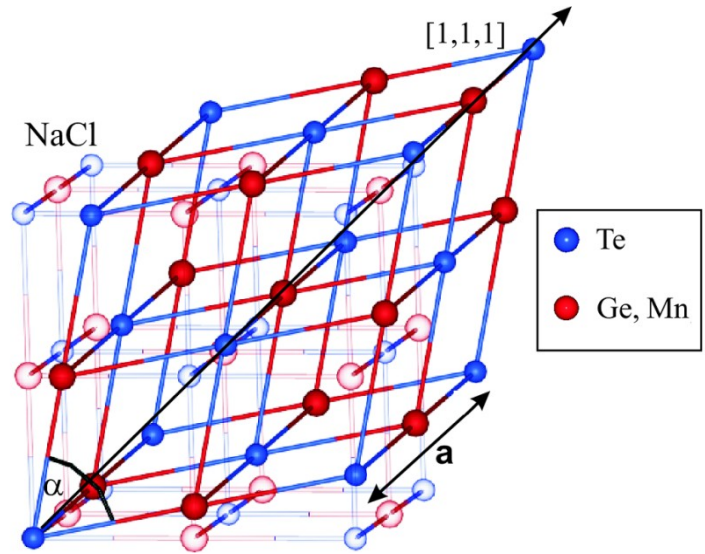

Figure 14. $Ge_{1-x}Mn_xTe$ crystal structure. Full markers – $0 \leq x \leq 0.18$, rhombohedral phase, ferroelectric. Empty markers – $0.18 \leq x \leq 0.60$, NaCl phase, paraelectric.

Multiferroic materials were investigated before either in heterostructures or in single phase compounds with a structure more complex than that of $Ge_{1-x}Mn_xTe$.[131-133] The most popular was $BiFeO_3$, also other perovskites.[134-136] Przybylińska *et al*. presented the first study of multiferroic properties of $Ge_{1-x}Mn_xTe$.[137] By measuring the ferromagnetic resonance the authors demonstrated that both types of order are coupled to each other. Their relative strength and properties can be tuned by changing the Mn content. The ferroelectric polarization may be switched by the external magnetic field. Besides, the ferromagnetic interactions may be controlled by the carrier concentration, as was shown above. These effects open a field for novel spintronics.

When the structural inversion symmetry is broken, the spin degeneracy is lifted and there appears a splitting of electronic states called Rashba splitting (for a review see, e.g. Reference [138]) Volobuev *et al*. described the giant Rasba splitting in $Pb_{1-x}Sn_xTe$ doped with bismuth.[139] The Rashba effect in GeTe and $Ge_{1-x}Mn_xTe$ was investigated and described by Krempaský *et al*. [140,141]

There were also theoretical and experimental studies on IV-VI DMS multiferroics based on SnTe.[142,143] The authors called them *triferroic*, i.e. ferroelectric + ferromagnetic + ferroelastic. The transition from the NaCl to the rhombohedral structure for bulk SnTe occurs at 100 K or below, depending on carrier concentration. However, in atomic-thick SnTe layer the transition temperature as high as 270 K was observed.

There are interesting theoretical *ab initio* calculations of the magnetic properties of GeTe based DMS. Łusakowski *et al.* calculated the effect of structural and magnetic disorder on the magnetic anisotropy in $Ge_{1-x}Mn_xTe$ layers.[144] Liu *et al*. presented calculations of exchange parameters in $Ge_{1-x}TM_xTe$ and $Sn_{1-x}TM_xTe$ where TM were V, Cr, and Mn.[145]

## 7.3. Heterostructures

MBE technique is especially useful for preparation of quantum well heterostructures. They were investigated with respect to their electrical and optical properties from the point of view of applications. As was mentioned above, Partin *et al*. prepared a whole series of heterostructures composed of PbTe and lead-rare-earth chalcogenides.[115–117,89] Such heterostructures could be applied as infrared diode lasers. Magnetotransport in $PbTe/Pb_{1-x}Eu_xTe$ quantum wells was investigated, spin alignment of electrons[146] and spin-orbit coupling were analysed.[147] Simma *et al*. measured optical properties of $PbSe/Pb_{1-x}Eu_xSe$ and $PbSe/Pb_{1-x}Eu_xSe_{1-y}Te_y$ quantum well heterostructures and investigated the band alignment in the structures.[148] Most of the research was focused on the transport and optical properties of the structures, not on the magnetic interactions.

Magnetic interactions were investigated in superlattices based on lead chalcogenides, built of magnetic (not diluted magnetic) epilayers intercalated with non-magnetic epilayers. Europium chalcogenides were often used as magnetic layers. Both lead chalcogenides and europium chalcogenides have the rocksalt structure. This is advantageous for MBE growth. A typical example of antiferromagnetic/non-magnetic structure is EuTe/PbTe. Heremans and Partin investigated magnetic properties of EuTe/PbTe superlattices and described the interlayer and intralayer magnetic interactions.[149] An example of ferromagnetic/non-magnetic structure is EuS/PbS, studied by Stachow-Wójcik et al.,[150] also described by Story.[151] Giebultowicz and Kępa compared these two structures in their studies of neutron scattering of interlayer magnetic coupling.[152] We will not describe here such structures, with very interesting magnetic properties, but not really DMS.

## 7.4. Nanocrystals

Now we may look for lower dimensional structures, such as nanowires and nanocrystals of IV-VI DMS. Quantum dots with magnetic ions are especially interesting as candidates for possible applications in spintronics and quantum information processing. They were investigated in II-VI DMS (see e.g. Ref. [153]) Up till now no systematic studies were performed on IV-VI DMS quantum dots. Optical properties of PbSe/Pb$_{1-x}$Eu$_x$Te structures, where non-magnetic PbSe dots were embedded in the diluted magnetic semiconductor lattice were investigated by Springholz et al. and Simma et al.[154,155] Dantas et al. prepared Pb$_{1-x}$Mn$_x$S nanocrystals embedded in glass, with sizes of about 6 nm and $0.003 \leq x \leq 0.01$.[156] Magnetization and magnetic susceptibility measurements have shown that the samples were paramagnetic with the Curie temperatures close to zero, like in the bulk materials. Silva et al. presented similar results.[157] Dantas et al. also produced nanocrystals of Pb$_{1-x}$Mn$_x$Se in glass by the same method.[158]

There is not much information in the literature about magnetic properties of IV-VI DMS quantum structures. They have so far not been considered for applications.

## 8. Topological crystalline insulators

In last ten years IV-VI mixed crystal semiconductors received new attention. This occurred due to the rapid development in research on topological insulators and introduction of a new topological state – a topological crystalline insulator (TCI). TCIs are topological insulators in which the role of the time-reversal symmetry is replaced by the crystal symmetry.[159, 160] It turned out that the IV-VI semiconductors may undergo a transition from the trivial insulator to the TCI phase. The transition is related to the band inversion which may be driven by a change in pressure, temperature, or crystal composition. The temperature-driven topological phase transition from a trivial insulator to a TCI in $Pb_{1-x}Sn_xSe$ was first reported by Dziawa *et al.*[161]

In IV-VI DMS we may observe an interplay of magnetic properties and energy band structure in TCIs. Before we analyze the influence of manganese on the band structure and topological properties of IV-VI mixed crystals we concentrate on the energy gap dependence on the composition. We report the results for $Pb_{1-x}Sn_xTe$ but the conclusions are valid for other mixed IV-VI crystals. For PbTe the wavefunctions of the valence band at the L point of the Brillouin zone are mainly built from *p* orbitals of anions while the conduction band is built from *p* orbitals of cations. The symmetries with respect to inversion are even and odd, what is denoted $L_6^+$ and $L_6^-$, respectively. In SnTe the situation is opposite, the valence band at the L point has symmetry $L_6^-$ and the conduction band has the symmetry $L_6^+$. Such situation is described as the inverted band structure and we say very often that SnTe has a negative energy gap. When we increase the tin content *x* we observe transition from the positive to the negative energy gaps. For $Pb_{1-x}Sn_xTe$ the transition occurs at *x* about 0.4, while for $Pb_{1-x}Sn_xSe$ it is at about 0.2. Ab initio calculations show that the transition is not sharp, i.e. it does not occur at the unique, well defined point. Instead there is certain nonzero width interval where the energy gap is zero. This is related to the chemical difference between two kinds of cations, i.e. chemical disorder. Due to this difference certain energy levels are split and as a consequence we observe a zero energy gap region. The longer discussion and more detailed explanation of this effect is presented in Reference [162]. Such a region of zero energy gap is also observed when, for a given *x,* we change the lattice parameter by application of hydrostatic pressure or by the change of temperature. The zero-energy gap region was experimentally confirmed,[163] although not by changing the composition but by applying the hydrostatic pressure. As was noticed by Wang *et al.* the zero energy gap region is the Weyl phase.

From experiment it is known that for $Pb_{1-x-y}Sn_xMn_yTe$ on PbTe side, i.e. for small *x*, addition of manganese leads to an increase in the energy gap.[86,164] *Ab initio* calculations are in agreement with this observation.[165] On SnTe side, i.e. for large *x*, the energy gap is negative and its absolute value decreases with increasing *x*. Therefore, the interval of zero energy gap region moves to higher values of tin content. Calculations of Spin Chern Numbers topological indices confirm that for small *x* the crystal is topologically trivial, while for large *x* it is nontrivial.[166] The width of the zero energy gap region, the Weyl phase, strongly depends on the magnetization of manganese ions. This fact may be understood by taking into account that in the presence of nonzero magnetization the band spin splitting always leads to a decrease of the modulus of the energy gap. That is why the Weyl phase appears at smaller *x* and disappears for larger *x* comparing to paramagnetic arrangement of manganese spins. This observation leads to a very interesting possibility of changing the topology of the bands by external magnetic field or the temperature.

The theoretical paper by Lusakowski *et al.* is devoted to topological properties of $Pb_{1-x-y}Sn_xMn_yTe$ bulk crystals.[165] There are numerous works in which the influence of magnetic ions on topological properties of surface states is analyzed both theoretically and experimentally.[167-170] In particular, quantum anomalous Hall effect (QAHE) is the object of many studies because it is important from the point of view of applications. Contrary to the quantum Hall effect, where very strong external magnetic fields are necessary to obtain edge modes, which in turn may be used to dissipationless transport of electric current, in QAHE such strong magnetic fields would not be necessary. There is hope that such edge modes may be obtained without external magnetic field, in the presence of magnetization due to magnetic ions. The main idea is that the exchange interaction between magnetic moments and free carriers can split the Dirac states localized on surfaces. This effect leads to the appearance of an energy gap. When the Fermi level lies in the gap we have a situation similar to the quantum Hall state. Indeed, recently QAHE was realized by Okazaki et al. by using topological insulator (Bi, Sb)Te doped with chromium and polarization of the magnetic ions was achieved by neodymium permanent magnet.[171] The applied magnetic field was much smaller than in the case of the quantum Hall effect.

In the context of IV-VI semiconductors this problem was analyzed by Fang *et al.*[172] The authors predicted possible Chern numbers up to the value of 4. They proposed also a heterostructure to realize QAHE. The experimental attempt to realize QAHE is described in Reference [173] where SnTe layers were doped with chromium. Another interesting experimental study is the work where the magnetic insulator EuS was placed near the the surface of SnTe crystal.[174] The magnetic field caused by EuS influences the surface topological states.

## 9. Summary

Diluted magnetic semiconductors based on IV-VI chalcogenides have not been investigated as extensively as those based on II-VI and III-V semiconducting compounds. They have, however, interesting magnetic properties. By looking at these compounds we may observe different types of magnetic interactions and phase transitions. There are short range interactions such as antiferromagnetic superexchange and long range interactions such as carrier mediated RKKY exchange. Moreover, by adjusting the carrier concentration we may observe *reversible* phase transitions from the paramagnetic phase (with small antiferromagnetic interactions) to the ferromagnetic phase. We may also observe phase transitions from paramagnetic to the spin glass phase, or paramagnetic-ferromagnetic-spin glass transitions.

The variety of magnetic interactions in IV-VI DMS leads sometimes to a complex character of the exchange mechanism and makes a precise description difficult. The reported values of exchange interactions in some rare-earth doped compounds with clusters or quaternary compounds show rather the general trend of behavior than the correct quantitative values. There is still no one model describing the magnetic specific heat in different IV-VI DMS, especially the large peak at zero field and the magnetic field dependence of the specific heat.

Curie temperatures up to 200 K, comparable with those found in III-V DMS, have been attained in IV-VI DMS thin films based on GeTe. This is prospective from the point of view of spintronics. There still remains a challenge to obtain the materials with Curie temperatures above the room temperature. More research is needed.

The GeTe-based compounds, because of the structural phase transition between the cubic and rhombohedral phase turned out to be the simplest multiferroic compounds. The possibilities of switching the ferroelectric polarization by the external magnetic field. and controlling the ferromagnetic interactions by the carrier concentration open a field for novel spintronics.

Recently, IV-VI mixed crystal semiconductors received new attention because of a rapid development in research on topological insulators. The IV-VI semiconductors may undergo a transition from the trivial insulator to the TCI phase. The transition is related to the band inversion which may be driven by a change in pressure, temperature, or crystal composition. The temperature-driven topological phase transition from a trivial insulator to a TCI was observed in IV-VI compounds. In recent years there are interesting studies on possibilities of realizing the quantum anomalous Hall effect in SnTe-based DMS.

The capability of tuning the energy band gap with the composition makes the IV-VI DMS useful in applications. A series of infrared detectors based on rare-earth doped compounds was successfully made. IV-VI DMS may be also used as infrared diodes and lasers. They also show some promise as thermoelectric materials. The applications, however, are not related to their magnetic properties.

Recently, there is some interesting research on nanocrystals of lead chalcogenides embedded in glasses or colloids. Nanostructures of IV-VI DMS need more investigation.

## Acknowledgements

This work was partially supported by National Science Centre NCN (Poland) projects
UMO-2016/23/B/ST3/03725 (AŁ), UMO-2017/27/B/ST3/02470 (AŁ), and
2018/30/E/ST3/00309.
We thank Professor Tomasz Story for helpful discussions.